\documentclass[fleqn,12pt]{article}
\usepackage[numbers,sort&compress]{natbib}
\usepackage[footnotesep=0.4in]{geometry}
\usepackage{graphicx}
\usepackage{amsmath}
\usepackage{bbding}
\usepackage{pifont}
\usepackage{amsfonts}
\usepackage{caption2}
\usepackage{float}
\usepackage{graphics}
\usepackage{mathrsfs}
\usepackage{amssymb}
\usepackage{booktabs}
\usepackage{tabularx}
\usepackage{multicol}
\usepackage{appendix}
\usepackage{graphicx}
\usepackage{epstopdf}
\usepackage{extarrows}
\graphicspath{{./pic/}} \makeatletter
\newcommand\figcaption{\def\@captype{figure}\caption}
\newcommand\tabcaption{\def\@captype{table}\caption}
\begin{document}
\date{}
\title{Painlev\'{e} analysis, B\"{a}cklund transformation, Lax pair and periodic wave solutions for a generalized (2+1)-dimensional Hirota-Satsuma-Ito equation in fluid mechanics}
\author{\hspace{-0.8cm}Dong Wang, Yi-Tian
Gao$^{}$\thanks{Corresponding author, with e-mail address as
gaoyt163@163.com}, Xin Yu$^{}$\thanks{Corresponding author, with e-mail address as
yuxin@buaa.edu.cn}, Gao-Fu Deng, Fei-Yan Liu\\
\\{\em Ministry-of-Education Key Laboratory of Fluid Mechanics and
National}\\
{\em  Laboratory for Computational Fluid Dynamics,  Beijing University of }\\
{\em  Aeronautics and Astronautics, Beijing 100191, China}}\maketitle
\newpage

\begin{abstract}
In this paper, we investigate a generalized (2+1)-dimensional Hirota-Satsuma-Ito (HSI) equation in fluid mechanics. Via the Painlev\'{e} analysis, we find that the HSI equation is Painlev\'{e} integrable under certain condition. Bilinear form, Bell-polynomial-type B\"{a}cklund transformation and Lax pair are constructed with the binary Bell polynomials. One-periodic-wave solutions are derived via the Hirota-Riemann method and displayed graphically.
\end{abstract}

 \vspace{20mm}
\newpage

\noindent{\Large{\bf 1. Introduction}}\vspace{5mm}

Nonlinear waves, including the lumps, solitons, periodic waves and rogue waves, have attracted researchers' interests in plasma physics, fluid mechanics and nonlinear optics~\cite{yusuf2021dynamics,baals2021stability,kruglov2021periodic}. For example, lumps generated by pressure disturbances moving over the liquid free surfaces have been investigated experimentally with the high-speed cameras and simulated numerically to study the wave patterns~\cite{masnadi2017observations,diorio2009gravity}. Optical solitons have been applied in the information transfer on transcontinental and transoceanic distances through optical fiber~\cite{khan2021two,houwe2021survey}. Propagation of periodic and solitary waves along the magnetic field has been examined in a cold collision-free plasma~\cite{abbas2020propagation}. Analyses of real-world ocean wave data and controlled experiments in wave tanks have supported both nonlinear and linear interpretations of rogue waves in the ocean~\cite{dudley2019rogue,cousins2019predicting,dematteis2019experimental}.

For the theoretical insights into those nonlinear waves, researchers have derived several nonlinear evolution equations (NLEEs) which have been verified experimentally, such as the nonlinear Shr\"{o}dinger equations, Kadomtsev-Petviashvili equations and Korteweg-de Vries equations~\cite{dematteis2019experimental,gao2019mathematical,mir2020forced,beji2018kadomtsev}.
To find the analytical solutions for those NLEEs, certain methods have been utilized, including the inverse scattering method, Hirota bilinear method, Darboux transformation, Kadomtsev-Petviashvili hierarchy reduction and Lie group analysis~\cite{chen2019conservation,chekhovskoy2021introducing,wazwaz2018new,du2020lie,zhang2020vector}.

References~\cite{ma2018study,kuo2020study,zhao2021bilinear,wang2021water} have investigated a generalized (2+1)-dimensional Hirota-Satsuma-Ito (HSI) equation in fluid mechanics, i.e.,
\begin{equation}\label{HSI}
u_{xxxt}+3 (u_x u_t)_x+\delta_1 u_{yt}+\delta_2 u_{xx}+\delta_3 u_{xy}+\delta_4 u_{xt}+\delta_5 u_{yy}=0,
\end{equation}
where $u=u(x,y,t)$ is a real function of the independent variables $x$, $y$ and $t$, the coefficients $\delta_\rho$'s ($\rho=1,2,\ldots,5$) are the real constants, and the subscripts with respect to $x$, $y$ and $t$ represent the partial derivatives. Special cases of Eq.~(\ref{HSI}) have been considered as follows:
\begin{itemize}
\item When $\delta_1=\delta_2=1$, $\delta_3=\delta_4=\delta_5=0$, Eq.~(\ref{HSI}) can be reduced to a (2+1)-dimensional HSI equation which describes the propagation of small-amplitude surface waves in a strait or large channels of slowly varying depth and width and non-vanishing vorticity~\cite{aliyu2020bell}.
\item When $\delta_2=\delta_4=-1$, $\delta_1=\delta_3=\delta_5=0$, under the transformation $q=u_x$, Eq.~(\ref{HSI}) can be reduced to the integrable Hirota-Satsuma shallow water wave equation~\cite{liu2019high}.
\end{itemize}

With the logarithm transformation $u=2(\ln f)_x$, the bilinear form for Eq.~(\ref{HSI}) has been obtained as~\cite{ma2018study}
\begin{equation}\label{Bilinear Form}
\left(D_x^3 D_t+\delta_1 D_y D_t+\delta_2 D_x^2+\delta_3 D_x D_y+\delta_4 D_x D_t+\delta_5 D_y^2 \right)f \cdot f=0,
\end{equation}
where $f=f(x,y,t)$ is a real differentiable function of $x$, $y$ and $t$, $D$ is the Hirota bilinear operator defined as~\cite{Hirota2004direct}
\begin{equation}\label{D-operator}
\begin{aligned}
&D^{A_1}_x D^{A_2}_y D^{A_3}_t G\cdot{F}=\left( \frac{\partial}{\partial x}-\frac{\partial}{\partial x'}\right) ^{A_1}
\left( \frac{\partial}{\partial y}-\frac{\partial}{\partial y'}\right) ^{A_2} \left( \frac{\partial}{\partial t}-\frac{\partial}{\partial t'}\right) ^{A_3} G(x,y,t)F(x',y',t')\Bigg| _{x'=x,~y'=y,~t'=t},
\end{aligned}
\end{equation}
with $G(x,y,t)$ as a function of $x$, $y$ and $t$, $F(x',y',t')$ as a function of the formal variables $x'$, $y'$ and $t'$, and $A_1$, $A_2$ and $A_3$ as the non-negative integers. Lump solutions for Eq.~(\ref{HSI}) have been obtained~\cite{ma2018study}. Through a direct computation, Ref.~\cite{kuo2020study} has illustrated that there is no resonant multi-soliton solution for Eq.~(\ref{HSI}). Bilinear B\"{a}cklund transformation (BT), kink and breather solutions for Eq.~(\ref{HSI}) have been studied~\cite{zhao2021bilinear}. Soliton, multiple-lump and hybrid solutions for Eq.~(\ref{HSI}) have been given~\cite{wang2021water}.

However, to our knowledge, Painlev\'{e} analysis, Bell-polynomial-type BT, Lax pair and periodic wave solutions for Eq.~(\ref{HSI}) have not been reported. In Sec. 2, Painlev\'{e} analysis for Eq.~(\ref{HSI}) will be worked out. In Sec. 3, via the binary Bell polynomials, bilinear form, Bell-polynomial-type BT and Lax pair for Eq.~(\ref{HSI}) will be obtained. In Sec. 4, the one-periodic wave solutions for Eq.~(\ref{HSI}) will be derived via the Hirota-Riemann method. In Sec. 5, the conclusions will be given.

\vspace{5mm} \noindent{\Large{\bf 2. Painlev\'{e} analysis for Eq.~(\ref{HSI}) }}\vspace{3mm}

Motivated by Ref.~\cite{ablowitz1977exact}, solutions for Eq.~(\ref{HSI}) can be expanded to the Laurent series,
\begin{equation}\label{Laurent}
\begin{aligned}
&u=\sum_{b=0}^{\infty} u_b \phi^{b+\alpha},
\end{aligned}
\end{equation}
where $\phi$ and $u_b$'s are the analytic functions of $x$, $y$ and $t$, $b$ is an integer, and $\alpha$ is a negative integer. The leading order of Expression~(\ref{Laurent}) can be assumed as
\begin{equation}\label{Laurent-leading}
\begin{aligned}
&u \sim u_0 \phi^\alpha,
\end{aligned}
\end{equation}
where $u_0$ is a nonzero function in the neighborhood of a non-characteristic movable singularity manifold.

Substituting Expression~(\ref{Laurent-leading}) into Eq.~(\ref{HSI}) and balancing the highest-order nonlinear and linear terms, we obtain $\alpha=-1$ and $u_0=2 \phi_x$. To find the resonance points, we can substitute
\begin{equation}\label{Laurent-leading-point}
\begin{aligned}
&u \sim u_0 \phi^{-1}+u_b \phi^{b-1}
\end{aligned}
\end{equation}
into Eq.~(\ref{HSI}) and let the sum of the terms with the lowest power of $\phi$ in Eq.~(\ref{HSI}) vanish, i.e.,
\begin{equation}\label{Laurent-leading-vanish}
\begin{aligned}
&(b^4-10 b^3+23 b^2+10 b-24) \phi^{-5+b} u_b \phi_x^3 \phi_t=0.
\end{aligned}
\end{equation}
Solving Eq.~(\ref{Laurent-leading-vanish}) yields four resonant points at $b=-1,1,4,6$.

Substituting
\begin{equation}\label{Laurent-compatibility}
\begin{aligned}
&u=\sum_{b=0}^{4} u_b \phi_{b-1}
\end{aligned}
\end{equation}
into Eq.~(\ref{HSI}) and setting the coefficients of $\phi^{b-5}$'s vanish, we can verify that the compatibility conditions are satisfied at the resonant points $b=1,4$, and that the corresponding $u_2$, $u_3$ and $u_5$ can be expressed. However, at the resonant point $b=6$, the compatibility condition is written as
\begin{equation}\label{compatibility condition}
\begin{aligned}
&6 \delta_5^2 \phi_x^2 h_1+3 \delta_1 \phi_t (2 \delta_5 \phi_y+\delta_3 \phi_x) h_2+2 \delta_5 h_3+\delta_3 h_4=0,
\end{aligned}
\end{equation}
where $h_\aleph=h_\aleph(u_1,u_4,\phi,\phi_x,\phi_y,\phi_t,\phi_{xx},\phi_{xy},\phi_{xt},\phi_{yt},\phi_{tt},\ldots)$ ($\aleph=1,2,3,4$) are the real functions relying on $\phi$. To satisfy Condition~(\ref{compatibility condition}) with whatever $\phi=\phi(x,y,t)$ takes, we have to set
\begin{equation}\label{compatibility condition-1}
\begin{aligned}
&\delta_3=\delta_5=0,
\end{aligned}
\end{equation}
and consequently Eq.~(\ref{HSI}) becomes Painlev\'{e} integrable.

It is worth noting that if we truncate Expression~(\ref{Laurent-compatibility}) into
\begin{equation}\label{Laurent-truncate}
\begin{aligned}
&u=u_0 \phi^{-1}+u_1,
\end{aligned}
\end{equation}
and set $u_1=0$, we obtain the aforementioned logarithm transformation $u=2(\ln f)_x$, which can be used to derive the bilinear form for Eq.~(\ref{HSI}).

\vspace{5mm} \noindent{\Large{\bf 3. Bell polynomials}}\vspace{3mm}

The multi-dimensional Bell polynomials are defined as~\cite{huang2017bilinear}
\begin{equation}\label{multi-dimensional-Bell-polynomials}
\begin{aligned}
&Y_{n_1 x_1,n_2 x_2,\ldots,n_l x_l}(\varphi)=Y_{n_1,n_2,\ldots,n_l}(\varphi_{r_1 x_1,r_2 x_2,\ldots,r_l x_l})=\mathrm{e}^{-\varphi} \partial_{x_1}^{n_1} \partial_{x_2}^{n_2} \cdots \partial_{x_l}^{n_l} \mathrm{e}^{\varphi},
\end{aligned}
\end{equation}
where $\varphi=\varphi(x_1,x_2,\ldots,x_l)$ is a $C^\infty$ function with multi-variables, $l$ is a nonnegative integer, $\varphi_{r_1 x_1,r_2 x_2,\ldots,r_l x_l}=\partial_{x_1}^{n_1} \partial_{x_2}^{n_2} \cdots \partial_{x_l}^{n_l} $, $n_1,n_2,\ldots,n_l$ are the nonnegative integers, and $r_1=0,1,\ldots,n_1$; $r_2=0,1,\ldots,n_2$; \ldots; $r_l=0,1,\ldots,n_l$. For instance, $Y_{x,t}=\varphi_{xt}+\varphi_x \varphi_t$, $Y_{2x,t}=\varphi_{xxt}+ \varphi_{xx}\varphi_t+2\varphi_{xt}\varphi_x+\varphi_x^2 \varphi_t$.

Based on Expression~(\ref{multi-dimensional-Bell-polynomials}), the binary Bell polynomials are written as~\cite{qin2017Bellpolynomial}
\begin{equation}\label{binary-Bell-polynomials}
\begin{aligned}
&\mathscr{Y}_{n_1 x_1,n_2 x_2,\ldots,n_l x_l}(v,w)=Y_{n_1,n_2,\ldots,n_l}(\varphi)\Big|_{\varphi_{r_1 x_1,r_2 x_2,\ldots,r_l x_l}}\\
&~~~~~~~~~~~~~~~~~~~~~~~~~~=\begin{cases}
v_{r_1 x_1,r_2 x_2,\ldots,r_l x_l} & \text{if $r_1+r_2+\cdots+r_l$ is odd},\\
w_{r_1 x_1,r_2 x_2,\ldots,r_l x_l} & \text{if $r_1+r_2+\cdots+r_l$ is even},
\end{cases}
\end{aligned}
\end{equation}
where $v$ and $w$ are the $C^\infty$ functions of $x_1,x_2,\ldots,x_l$. For example, $\mathscr{Y}_x(v,w)=v_x,~\mathscr{Y}_{2x}(v,w)=w_{xx}+v_x^2,~\mathscr{Y}_{x,t}(v,w)=v_x v_t+w_{xt},~\mathscr{Y}_{3x}(v,w)=v_{xxx}+3v_x w_{xx}+v_x^3$. The relationship between the $n_1+n_2+\cdots+n_l$th order $\mathscr{Y}$-polynomials and the Hirota bilinear operators is expressed as~\cite{tian2010Riemann}
\begin{equation}\label{binary-Bell-polynomials-relation}
\begin{aligned}
&\mathscr{Y}_{n_1 x_1,n_2 x_2,\ldots,n_l x_l}[v=\ln (G/F),w=\ln (GF)]=(G \cdot F)^{-1} D_{x_1}^{n_1} D_{x_2}^{n_2} \cdots D_{x_l}^{n_l} G \cdot F,
\end{aligned}
\end{equation}
where $n_1+n_2+\cdots+n_l \geq 1$, $G$ and $F$ are the $C^\infty$ functions of $x_1,x_2,\ldots,x_l$.

In the particular case of $F=G$, Eq.~(\ref{binary-Bell-polynomials-relation}) can be written as
\begin{equation}\label{binary-Bell-polynomials-relation-1}
\begin{aligned}
&F^{-2}D_{x_1}^{n_1} D_{x_2}^{n_2} \cdots D_{x_l}^{n_l}F \cdot F=\mathscr{Y}_{n_1 x_1,n_2 x_2,\ldots,n_l x_l}(0,q=2\ln F).
\end{aligned}
\end{equation}
Examples for the $P$-polynomials are shown as:
\begin{equation}\label{binary-Bell-example}
\begin{aligned}
&P_{2x}(q)=q_{xx},~P_{x,t}(q)=q_{xt},~P_{4x}(q)=q_{xxxx}+3q_{xx}^2, ~P_{3x,y}(q)=q_{xxxy}+3q_{xx}q_{xy}.
\end{aligned}
\end{equation}

$\mathscr{Y}$-polynomials can be separated into $P$-polynomials and $Y$-polynomials as~\cite{huang2017bilinear}
\begin{equation}\label{binary-Bell-separate}
\begin{aligned}
&~~~~(G \cdot F)^{-1}D_{x_1}^{n_1} D_{x_2}^{n_2} \cdots D_{x_l}^{n_l} G \cdot F\\
&=\mathscr{Y}_{n_1 x_1,n_2 x_2,\ldots,n_l x_l}(v,w)\Big|_{v=\ln (G/F),~w=\ln (GF)}\\
&=\mathscr{Y}_{n_1 x_1,n_2 x_2,\ldots,n_l x_l}(v,v+q)\Big|_{v=\ln (G/F),~q=2\ln F}\\
&=\sum_{n_1+n_2+\cdots+n_l=\text{even}} \sum_{r_1=0}^{n_1} \sum_{r_2=0}^{n_2} \cdots \sum_{r_l=0}^{n_l} \prod_{\imath=q}^l \begin{pmatrix} n_\imath \\ r_\imath \end{pmatrix} P_{r_1 x_1,r_2 x_2,\ldots,r_l x_l} Y_{(n_1-r_1)x_1,(n_2-r_2)x_2,\ldots,(n_l-r_l)x_l}(v).
\end{aligned}
\end{equation}

One property of the multi-dimensional Bell polynomials, i.e.,~\cite{qin2017Bellpolynomial}
\begin{equation}\label{binary-Bell-separate-y}
\begin{aligned}
&\mathscr{Y}_{(n_1-r_1)x_1,(n_2-r_2)x_2,\ldots,(n_l-r_l)x_l}(v)\Big|_{v=\ln \psi}=\frac{\psi_{(n_1-r_1)x_1,(n_2-r_2)x_2,\ldots,(n_l-r_l)x_l}}{\psi},
\end{aligned}
\end{equation}
implies that the $\mathscr{Y}$-polynomials can be linearized by means of the Cole-Hopf transformation $v=\ln \psi$, where $\psi$ is a real function of $x_1,x_2,\ldots,x_l$.

\vspace{5mm}\noindent{\large{\bf 3.1. Bilinear form for Eq.~(\ref{HSI})}}\vspace{5mm}

In order to seek the linearizable expression of Eq.~(\ref{HSI}), we assume
\begin{equation}\label{binary-Bell-u-assume}
\begin{aligned}
&u=c\bar{q}_x,
\end{aligned}
\end{equation}
where $c$ is a real constant, $\bar{q}$ is a real function of $x$, $y$ and $t$. Substituting Expression~(\ref{binary-Bell-u-assume}) into Eq.~(\ref{HSI}) and integrating the equation with respect to $x$, we have
\begin{equation}\label{binary-Bell-u-eq}
\begin{aligned}
&\bar{q}_{xxxt}+3c\bar{q}_{xx}\bar{q}_{xt}+\delta_1 \bar{q}_{yt}+\delta_2 \bar{q}_{xx}+\delta_3 \bar{q}_{xy}+\delta_4 \bar{q}_{xt}+\delta_5 \bar{q}_{yy}=\epsilon,
\end{aligned}
\end{equation}
where $\epsilon$ is a real constant. Introducing Expression~(\ref{binary-Bell-example}) and setting $c=1$, Eq.~(\ref{binary-Bell-u-eq}) can be transformed into the $P$-polynomials as
\begin{equation}\label{binary-Bell-u-eq1}
\begin{aligned}
&P_{3xt}(\bar{q})+\delta_1 P_{yt}(\bar{q})+\delta_2 P_{2x}(\bar{q})+\delta_3 P_{xy}(\bar{q}) +\delta_4 P_{xt}(\bar{q})+\delta_5 P_{2y}(\bar{q})=\epsilon.
\end{aligned}
\end{equation}
Considering Expression~(\ref{binary-Bell-polynomials-relation-1}) and setting
\begin{equation}\label{binary-Bell-q-set}
\begin{aligned}
&\bar{q}=2\ln f,~\epsilon=0,
\end{aligned}
\end{equation}
we obtain the bilinear form of Eq.~(\ref{HSI}) as
\begin{equation}\label{Bilinear Form-1}
\left(D_x^3 D_t+\delta_1 D_y D_t+\delta_2 D_x^2+\delta_3 D_x D_y+\delta_4 D_x D_t+\delta_5 D_y^2 \right)f \cdot f=0,
\end{equation}
which is in accord with the one in Reference~\cite{ma2018study}.

\vspace{5mm}\noindent{\large{\bf 3.2. Bell-polynomial-type BT and Lax pair for Eq.~(\ref{HSI})}}\vspace{5mm}

In what follows, we will derive the Bell-polynomial-type BT and Lax pair for Eq.~(\ref{HSI}). Supposing that
\begin{equation}\label{binary-Bell-q-suppose}
\hat{q}=2\ln F,~\tilde{q}=2\ln G,
\end{equation}
where $u=\hat{q}_x$ and $u=\tilde{q}_x$ are different solutions for Eq.~(\ref{HSI}), $\hat{q}$ and $\tilde{q}$ are real functions of $x$, $y$ and $t$. Setting
\begin{equation}\label{binary-Bell-wv-suppose}
v=\frac{\tilde{q}-\hat{q}}{2}=\ln \frac{G}{F},~w=\frac{\tilde{q}+\hat{q}}{2}=\ln (GF),
\end{equation}
we can obtain a condition for the Bell-polynomial-type BT as
\begin{equation}\label{binary-Bell-e-suppose}
E(\tilde{q})-E(\hat{q})=0,
\end{equation}
where
\begin{equation}\label{binary-Bell-e-expand}
\begin{aligned}
&~~~~E(\tilde{q})-E(\hat{q})\\
&=[P_{3xt}(w+v)-P_{3xt}(w-v)]+\delta_1 [P_{yt}(w+v)-P_{yt}(w-v)]+\delta_2 [P_{2x}(w+v)-P_{2x}(w-v)]\\
&~~+\delta_3 [P_{xy}(w+v)-P_{xy}(w-v)]+\delta_4 [P_{xt}(w+v)-P_{xt}(w-v)]+\delta_5 [P_{2y}(w+v)-P_{2y}(w-v)]\\
&=2(v_{xxxy}+3w_{xx}v_{xy}+3v_{xx}w_{xy}+\delta_1 v_{yt}+\delta_2 v_{xx}+\delta_3 v_{xy}+\delta_4 v_{xt}+\delta_5 v_{yy})\\
&=2\partial_y(v_{xxx}+3v_x w_{xx}+v_x^3)-2\partial_y (3v_x w_{xx}+v_x^3)+6(w_{xx} v_{xy}+v_{xx}w_{xy})\\
&~~+2\partial_y (\delta_1 v_t+\delta_3 v_x+\delta_5 v_y)+2\partial_x (\delta_2 v_x+\delta_4 v_t)\\
&=2\partial_y(v_{xxx}+3v_x w_{xx}+v_x^3+\delta_1 v_t+\delta_3 v_x+\delta_5 v_y)+2\partial_x (\delta_2 v_x+\delta_4 v_t)\\
&~~-2\partial_y (3v_x w_{xx}+v_x^3)+6(w_{xx} v_{xy}+v_{xx} w_{xy}),
\end{aligned}
\end{equation}

Introducing the constraint equation,
\begin{equation}\label{binary-Bell-e-constraint-equation}
\begin{aligned}
&w_{xy}+v_x v_y=\lambda(t),
\end{aligned}
\end{equation}
where $\lambda(t)$ is a real function of $t$, we can covert Expression~(\ref{binary-Bell-e-expand}) to
\begin{equation}\label{binary-Bell-e-equation}
\begin{aligned}
&E(\tilde{q})-E(\hat{q})=2\partial_y \left[\mathscr{Y}_{3x}(v,w)+\delta_1 \mathscr{Y}_t (v,w)+\delta_3 \mathscr{Y}_x(v,w)+\delta_5 \mathscr{Y}_y(v,w) \right]\\
&~~~~~~~~~~~~~~~~~~~+2\partial_x \left[3\lambda(t) \mathscr{Y}_x(v,w)+\delta_2 \mathscr{Y}_x(v,w)+\delta_4 \mathscr{Y}_t(v,w) \right].
\end{aligned}
\end{equation}
Therefore, Eq.~(\ref{binary-Bell-e-suppose}) can be transformed into
\begin{equation}\label{binary-Bell-e-equation}
\begin{aligned}
&\begin{cases}
\mathscr{Y}_{3x}(v,w)+\delta_1 \mathscr{Y}_t(v,w)+\delta_3 \mathscr{Y}_x(v,w)+\delta_5 \mathscr{Y}_y(v,w)=0,\\
3\lambda (t) \mathscr{Y}_x(v,w)+\delta_2 \mathscr{Y}_x(v,w)+\delta_4 \mathscr{Y}_t(v,w)=0,\\
\mathscr{Y}_{xy}(v,w)-\lambda(t)=0.
\end{cases}
\end{aligned}
\end{equation}

Hereby, we obtain the Bell-polynomial-type BT for Eq.~(\ref{HSI}) as
\begin{equation}\label{Bell-polynomial-type-BT}
\begin{aligned}
&\begin{cases}
\left( D_x^3+\delta_1 D_t+\delta_3 D_x+\delta_5 D_y \right)G \cdot F=0,\\
\left[ (3\lambda+\delta_2) D_x+\delta_4 D_t \right]G \cdot F=0,\\
\left( D_x D_y-\lambda \right)G \cdot F=0,
\end{cases}
\end{aligned}
\end{equation}
which is different from the BT derived in Reference~\cite{zhao2021bilinear}.

With $v=\ln \psi$, Eq.~(\ref{binary-Bell-separate}) can be transferred to
\begin{equation}\label{binary-Bell-separate-y}
\begin{aligned}
&\mathscr{Y}_{x}(v,w)=\frac{\psi_x}{\psi},~\mathscr{Y}_y(v,w)=\frac{\psi_y}{\psi}, ~\mathscr{Y}_t(v,w)=\frac{\psi_t}{\psi},~\mathscr{Y}_{xy}(v,w)=q_{xy}+\frac{\psi_{xy}}{\psi},\\ &\mathscr{Y}_{3x}(v,w)=\frac{\psi_{xxx}}{\psi}+\frac{3q_{xx} \psi_x}{\psi},~\mathscr{Y}_{2x,y}=\frac{q_{2x} \psi_y}{\psi}+\frac{2q_{xy} \psi_x}{\psi}+\frac{\psi_{xxy}}{\psi}.
\end{aligned}
\end{equation}

Substituting $v=\ln \psi$ into Eqs.~(\ref{binary-Bell-e-equation}), and then applying Eq.~(\ref{binary-Bell-separate-y}), we can obtain the Lax pair for Eq.~(\ref{HSI}) as
\begin{equation}\label{Bell-polynomial-Lax-pair}
\begin{aligned}
&\begin{cases}
\psi_{xxx}+3q_{xx} \psi_x+\delta_1 \psi_t+\delta_3 \psi_x+\delta_5 \psi_y=0,\\
(3\lambda+\delta_2) \psi_x +\delta_4 \psi_t=0,\\
q_{xy} \psi+\psi_{xy}-\lambda \psi=0.
\end{cases}
\end{aligned}
\end{equation}

\vspace{5mm} \noindent{\Large{\bf 4. Periodic wave solutions for Eq.~(\ref{HSI})}}\vspace{3mm}

In order to study the one-periodic wave solutions for Eq.~(\ref{HSI}), we introduce the one-Riemann theta function as~\cite{shen2021breather}
\begin{equation}\label{one-Riemann-theta-function}
\begin{aligned}
&\theta(\xi,\tau)=\sum_{n=-\infty}^{+\infty} \mathrm{e}^{\mathrm{i} \pi n^2 \tau+2 \mathrm{i} \pi n \xi},
\end{aligned}
\end{equation}
where $n$ is an integer, $\mathrm{i}=\sqrt{-1}$, $\xi=\mu x+\nu y+\gamma t+\zeta$, $\tau$ is a complex constant satisfying $\text{Im} (\tau) >0$, and $\mu$, $\nu$, $\gamma$ and $\zeta$ are the real constants. Substituting Expression~(\ref{one-Riemann-theta-function}) into Bilinear Form~(\ref{Bilinear Form}), we have
\begin{equation}\label{Riemann-theta-1}
\begin{aligned}
&\bar{G}[D_x,D_y,D_t,c]\theta (\xi,\tau) \cdot \theta (\xi,\tau) \triangleq [ D_x^3 D_t+\delta_1 D_y D_t+\delta_2 D_x^2+\delta_3 D_x D_y\\
&~~~~~~~~~~~~~~~~~~~~~~~~~~~~~~~~~~~~~~~~~~+\delta_4 D_x D_t+\delta_5 D_y^2+c ] \theta (\xi,\tau) \cdot \theta (\xi,\tau)=0,
\end{aligned}
\end{equation}
where $c$ is a constant to be known later.

With the following property of the Hitora bilinear operators,
\begin{equation}\label{Hirota property}
\begin{aligned}
D_x^{A_1} D_y^{A_2} D_t^{A_3} \mathrm{e}^{\xi_1}\cdot \mathrm{e}^{\xi_2}=(\mu_1-\mu_2)^{A_1} (\nu_1-\nu_2)^{A_2} (\gamma_1-\gamma_2)^{A_3} \mathrm{e}^{\xi_1+\xi_2},
\end{aligned}
\end{equation}
where $\mu_1$, $\mu_2$, $\nu_1$, $\nu_2$, $\gamma_1$, $\gamma_2$, $\zeta_1$ and $\zeta_2$ are the real constants, $\xi_1=\mu_1 x+\nu_1 y+\gamma_1 t+\zeta_1$, and $\xi_2=\mu_2 x+\nu_2 y+\gamma_2 t+\zeta_2$, Expression~(\ref{Riemann-theta-1}) can be derived as
\begin{equation}\label{Riemann-theta-2}
\begin{aligned}
&~~~~\bar{G}[D_x,D_y,D_t,c]\theta (\xi,\tau) \cdot \theta (\xi,\tau)\\
&=[D_x^3 D_t+\delta_1 D_y D_t+\delta_2 D_x^2+\delta_3 D_x D_y+\delta_4 D_x D_t+\delta_5 D_y^2+c] \sum_{n=-\infty}^{+\infty} \mathrm{e}^{\mathrm{i} \pi n^2 \tau+2 \mathrm{i} \pi n \xi} \cdot \sum_{m=-\infty}^{+\infty} \mathrm{e}^{\mathrm{i} \pi m^2 \tau+2 \mathrm{i} \pi m \xi}\\
&=\sum_{n=-\infty}^{+\infty} \sum_{m=-\infty}^{+\infty} [D_x^3 D_t+\delta_1 D_y D_t+\delta_2 D_x^2+\delta_3 D_x D_y+\delta_4 D_x D_t+\delta_5 D_y^2+c] \mathrm{e}^{\mathrm{i} \pi n^2 \tau+2 \mathrm{i} \pi n \xi} \cdot \mathrm{e}^{\mathrm{i} \pi m^2 \tau+2 \mathrm{i} \pi m \xi}\\
&=\sum_{n=-\infty}^{+\infty} \sum_{m=-\infty}^{+\infty} \bar{G} [2 \mathrm{i} \pi (n-m) \mu,2 \mathrm{i} \pi (n-m) \nu,2 \mathrm{i} \pi (n-m) \gamma] \mathrm{e}^{2\mathrm{i} \pi (n+m) \xi +\mathrm{i} \pi (n^2+m^2) \tau}\\
&\xlongequal[m^\prime=n+m]{def}\sum_{m^\prime=-\infty}^{+\infty} \hat{G}[m^\prime] \mathrm{e}^{2\mathrm{i}\mathrm{\pi} m^\prime \xi},
\end{aligned}
\end{equation}
where $m$ is an integer and
\begin{equation}\label{Riemann-theta-3}
\begin{aligned}
&\hat{G}[m^\prime] \triangleq \sum_{n=-\infty}^{+\infty} \bar{G} [2 \mathrm{i} \pi (2n-m^\prime) \mu,2 \mathrm{i} \pi (2n-m^\prime) \nu,2 \mathrm{i} \pi (2n-m^\prime) \gamma] \mathrm{e}^{\mathrm{i} \pi [n^2+(n-m^\prime)^2] \tau}\\
&~~~~~~~\xlongequal[n^\prime=n-1]{def} \sum_{n^\prime=-\infty}^{+\infty} \bar{G} [2 \mathrm{i} \pi [2n^\prime-(m^\prime-2)] \mu,[2 \mathrm{i} \pi [2n^\prime-(m^\prime-2)] \nu,[2 \mathrm{i} \pi [2n^\prime-(m^\prime-2)] \gamma] \\
&~~~~~~~~~~~~~~~~~~\cdot \mathrm{e}^{\mathrm{i} \pi [n^2-(m^\prime-2)^2] \tau} \cdot \mathrm{e}^{2 \mathrm{i} \pi (m^\prime-1)\tau}\\
&~~~~~~~=\hat{G}[m^\prime -2] \mathrm{e}^{2 \mathrm{i} \pi (m^\prime-1) \tau}\\
&~~~~~~~=\ldots=\begin{cases}
\hat{G}[0] \mathrm{e}^{\frac{1}{2}m^{\prime2} \mathrm{i}\mathrm{\pi} \tau},~~~~~(m^\prime~\text{is even}),\\
\hat{G}[1] \mathrm{e}^{\frac{1}{2}(m^{\prime2}-1) \mathrm{i}\mathrm{\pi} \tau},~(m^\prime~\text{is odd}).
                                \end{cases}
\end{aligned}
\end{equation}

To ensure that Expression~(\ref{Riemann-theta-2}) satisfies Eq.~(\ref{Riemann-theta-1}), we set $\hat{G}[0]=0$ and $\hat{G}[1]=0$, where
\begin{equation}\label{Riemann-G}
\begin{aligned}
&\hat{G}[0]=\sum_{n=-\infty}^{+\infty} \bar{G}[2\mathrm{i}\mathrm{\pi} (2n)\mu,2\mathrm{i}\mathrm{\pi} (2n)\nu,2\mathrm{i}\mathrm{\pi} (2n)\gamma] \mathrm{e}^{2\mathrm{i}\mathrm{\pi} n^2 \tau}\\
&~~~~~~=\sum_{n=-\infty}^{+\infty} (256\pi^4 n^4 \mu^3 \gamma-16 \delta_1 \pi^2 n^2 \nu \gamma-16 \delta_2 \pi^2 n^2 \mu^2-16 \delta_3 \pi^2 n^2 \mu \nu-16 \delta_4 \pi^2 n^2 \mu \gamma\\
&~~~~~~~~~-16 \delta_5 \pi^2 n^2 \nu^2+c) \mathrm{e}^{2\mathrm{i}\mathrm{\pi} n^2 \tau},\\
&\hat{G}[1]=\sum_{n=-\infty}^{+\infty} \bar{G}[2\mathrm{i}\mathrm{\pi} (2n-1)\mu,2\mathrm{i}\mathrm{\pi} (2n-1)\nu,2\mathrm{i}\mathrm{\pi} (2n-1)\gamma] \mathrm{e}^{\mathrm{i} \pi (2n^2-2n+1)\tau}\\
&~~~~~~=\sum_{n=-\infty}^{+\infty} [16 \pi^4 (2n-1)^4 \mu^3\gamma-4\delta_1 \pi^2 (2n-1)^2 \nu\gamma-4 \delta_2 \pi^2 (2n-1)^2 \mu^2\\
&~~~~~~~~~~-4 \delta^3 \pi^2 (2n-1)^2 \mu\nu-4 \delta_4 \pi_2 (2n-1)^2 \mu\gamma-4 \delta_5 \pi_2 (2n-1)^2 \nu^2+c] \mathrm{e}^{\mathrm{i} \pi (2n^2-2n+1)\tau}.
\end{aligned}
\end{equation}
Therefore, $\hat{G}[0]=0$ and $\hat{G}[1]=0$ are equivalent to
\begin{equation}\label{Matrix-abc}
\begin{aligned}
  \begin{bmatrix}a_{11} & a_{12}\\a_{21} & a_{22}\end{bmatrix}
  \begin{bmatrix}\gamma\\c\end{bmatrix}=
  \begin{bmatrix}b_1\\b_2\end{bmatrix},
\end{aligned}
\end{equation}
where
\begin{equation}\label{Matrix-elements}
\begin{aligned}
&a_{11}=\sum_{n=-\infty}^{+\infty} (256 \pi^4 n^4 \mu^3-16 \delta_1 \pi^2 n^2 \nu-16 \delta_4 \pi^2 (2n-1)^2 \mu\gamma) \mathrm{e}^{2\mathrm{i}\mathrm{\pi} n^2 \tau},\\
&a_{21}=\sum_{n=-\infty}^{+\infty} [16 \pi^4 (2n-1)^4 \mu^3-4 \delta_1 \pi^2 (2n-1)^2 \nu-4 \delta_4 \pi^2 (2n-1)^2 \mu] \mathrm{e}^{\mathrm{i} \pi (2n^2-2n+1)\tau},\\
&a_{12}=\sum_{n=-\infty}^{+\infty} \mathrm{e}^{2\mathrm{i}\mathrm{\pi} n^2 \tau},~a_{22}=\sum_{n=-\infty}^{+\infty}\mathrm{e}^{\mathrm{i} \pi (2n^2-2n+1)\tau},\\
&b_1=\sum_{n=-\infty}^{+\infty} (16 \delta_2 \pi^2 n^2 \mu^2+16 \delta_3 \pi^2 n^2 \mu\nu+16 \delta_5 \pi^2 n^2 \nu^2) \mathrm{e}^{2\mathrm{i}\mathrm{\pi} n^2 \tau},\\
&b_2=\sum_{n=-\infty}^{+\infty} [4 \delta_2 \pi^2 (2n-1)^2 \mu^2+4\delta_3 \pi^2 (2n-1)^2\mu\nu+4\delta_5 \pi^2 (2n-1)^2 \nu^2] \mathrm{e}^{\mathrm{i} \pi (2n^2-2n+1)\tau}.
\end{aligned}
\end{equation}
Solving Expression~(\ref{Matrix-abc}), we can obtain
\begin{equation}\label{Matrix-solution}
\begin{aligned}
&\gamma=\frac{a_{22} b_1-a_{12} b_2}{a_{11} a_{22}-a_{12} a_{21}},~c=\frac{a_{21} b_1-a_{11} b_2}{a_{12} a_{21}-a_{11} a_{22}}.
\end{aligned}
\end{equation}
According to the theorems in Ref.~\cite{shen2021breather}, the one-periodic wave solutions for Eq.~(\ref{HSI}) can be expressed as
\begin{equation}\label{one-periodic-solution}
\begin{aligned}
u=u_0+2 \left[\ln \theta(\xi,\tau) \right]_x,
\end{aligned}
\end{equation}
where $u_0$ is a real constant satisfying the asymptotic
condition $u \rightarrow u_0$ when $\vert \xi \vert \rightarrow 0$. Figs. 1 present the propagation of one-periodic wave on the $x-y$ plane with $\delta_\rho=1$, $\mu=\nu=0.5$, $\tau=0.5\mathrm{i}$ and $\zeta=0$. In Figs. 1, we can observe that the wave profile of one-periodic wave is unchanged during the propagation.

\vspace{5mm}\noindent\begin{minipage}{\textwidth}
\renewcommand{\captionfont}{\scriptsize}
\renewcommand{\captionlabelfont}{\scriptsize}
\renewcommand{\captionlabeldelim}{.\,}
\renewcommand{\figurename}{Figs.\,}
\center\includegraphics[scale=0.5]{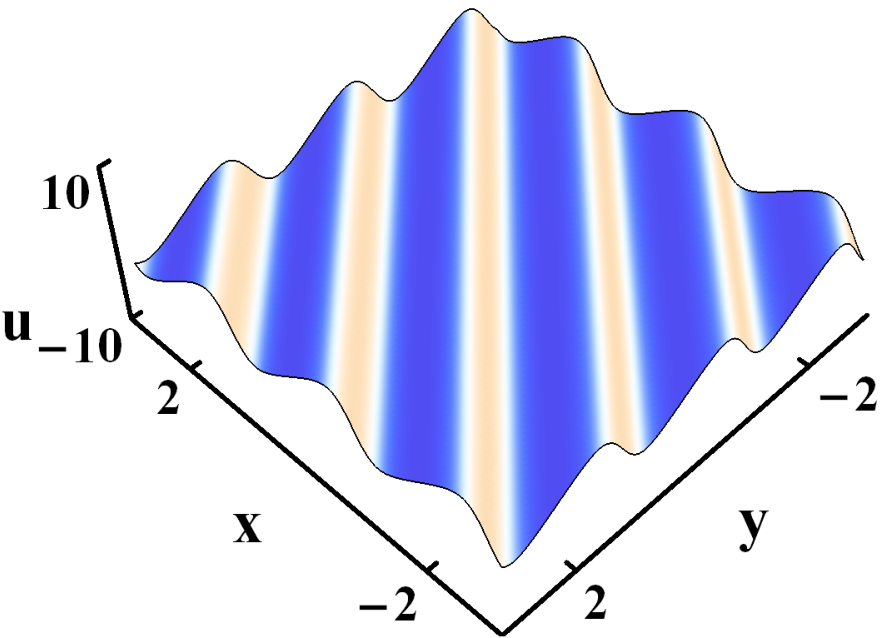}\hspace{4mm}\includegraphics[scale=0.5]{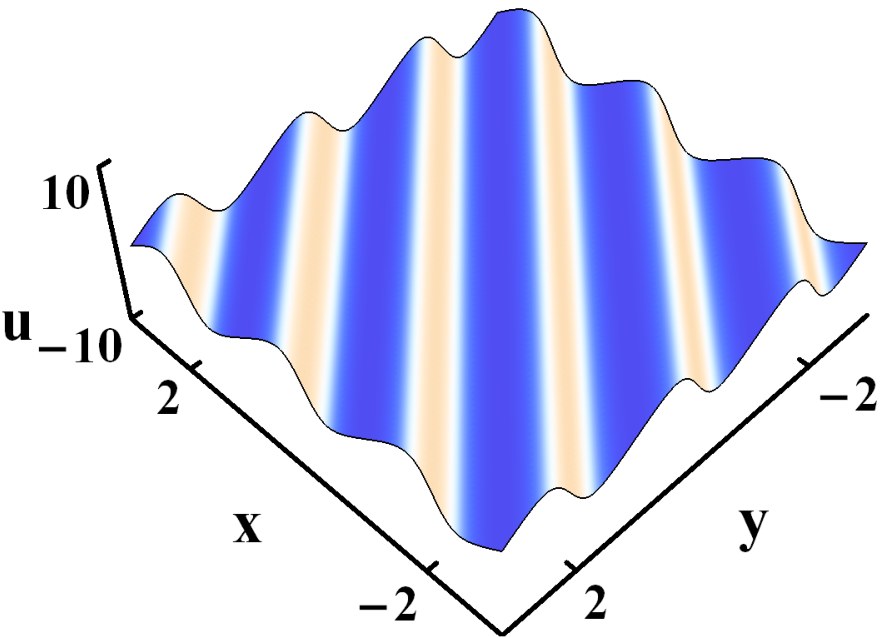}
\hspace{4mm}\includegraphics[scale=0.5]{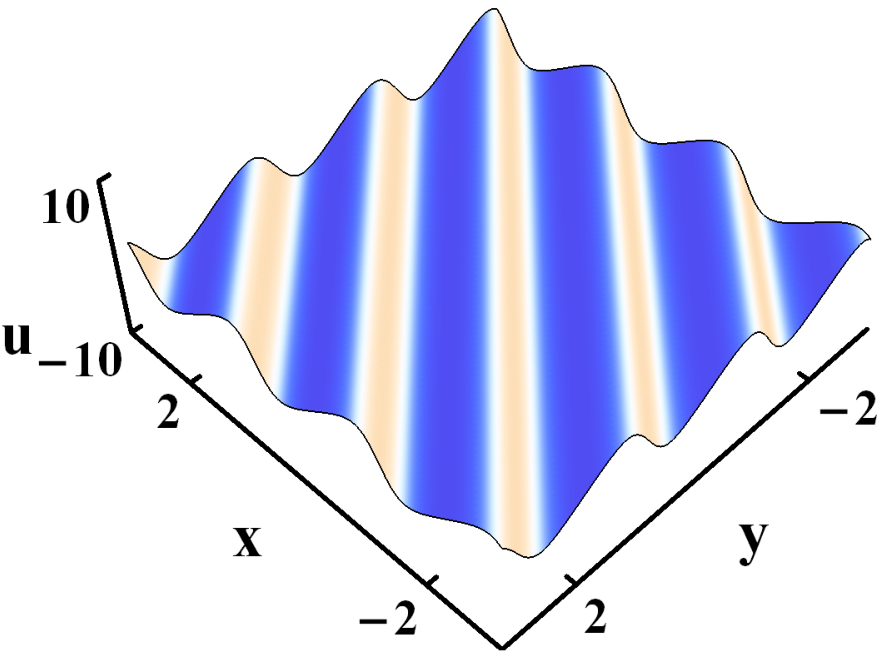}\\
{\center\hspace{0.1cm}\footnotesize(a)~$t=-1$ \hspace{3.5cm}(b) ~$t=0$\hspace{3.5cm}(c) ~$t=1$}\\
\noindent\figcaption{The one-periodic wave via Solutions~(\ref{one-periodic-solution}), with
$\delta_\rho=1$, $\mu=\nu=0.5$, $\tau=0.5\mathrm{i}$ and $\zeta=0$.
}
\label{one-periodic-fig1}
\end{minipage}\vspace{5mm}

\vspace{5mm} \noindent{\Large{\bf 6. Conclusions}}\vspace{3mm}

In this paper, we have investigated a generalized (2+1)-dimensional HSI equation in fluid mechanics, i.e., Eq.~(\ref{HSI}). Via Painlev\'{e} analysis, we have found that Eq.~(\ref{HSI}) is Painlev\'{e} integrable under Condition~(\ref{compatibility condition-1}). By truncating Laurent Series~(\ref{Laurent}), we have obtained the logarithm transformation $u=2(\ln f)_x$, which can be used to derive the bilinear form for Eq.~(\ref{HSI}).

Via the binary Bell polynomials, we have derived the bilinear form, Bell-polynomial-type BT, Lax pair for Eq.~(\ref{HSI}), i.e., Eqs.~(\ref{Bilinear Form-1}), (\ref{Bell-polynomial-type-BT}) and (\ref{Bell-polynomial-Lax-pair}), respectively. By virtue of the Hirota-Riemann method, we have constructed the one-periodic wave solutions for Eq.~(\ref{HSI}), i.e., Solutions~(\ref{one-periodic-solution}). Figs. 1 have shown the propagation of the one-periodic wave on the $x-y$ plane.

\vspace{5mm}\noindent{\Large{\bf Acknowledgement}}\vspace{3mm}

We express our sincere thanks to all the members of our discussion
group for their valuable comments. This work has been supported by the National Natural Science Foundation of China under Grant No. 11272023, and by the Fundamental Research Funds for the Central Universities.

\end{document}